# Citation analysis cannot legitimate the strategic selection of excellence

Tobias Opthof [1,2] & Loet Leydesdorff [3]

**Abstract**

In reaction to a previous critique (Opthof & Leydesdorff, 2010), the Center for Science and Technology Studies (CWTS) in Leiden proposed to change their old "crown" indicator in citation analysis into a new one. Waltman *et al*. (2011a) argue that this change does not affect rankings at various *aggregated* levels. However, CWTS data is not publicly available for testing and criticism. In this correspondence, we use previously published data of Van Raan (2006) to address the pivotal issue of how the results of citation analysis correlate with the results of peer review. A quality parameter based on peer review was neither significantly correlated with the two parameters developed by the CWTS in the past (CPP/JCS$_m$ or CPP/FCS$_m$) nor with the more recently proposed *h*-index (Hirsch, 2005). Given the high correlations between the old and new "crown" indicators, one can expect that the lack of correlation with the peer-review based quality indicator applies equally to the newly developed ones.

---

[1] Experimental Cardiology Group, Heart Failure Research Center, Academic Medical Center AMC, Meibergdreef 9, 1105 AZ Amsterdam, The Netherlands; t.opthof@inter.nl.net.
[2] Department Of Medical Physiology, University Medical Center Utrecht, The Netherlands
[3] Amsterdam School of Communication Research (ASCoR), University of Amsterdam, Kloveniersburgwal 48, 1012 CX Amsterdam, The Netherlands; loet@leydesdorff.net .

In this Letter, we react on a study by Waltman *et al.* (2011a), entitled "Towards a new crown indicator: An empirical analysis." These authors go at great length to show that a change in the normalization—in reaction to our previous critique of the Leiden "crown" indicators (Opthof & Leydesdorff, 2010)—did not significantly affect the rankings at various *aggregated* levels. Since the CWTS-data under discussion are not publicly available, we use a previous occasion at which Van Raan (2006) revealed some of the micro-data underlying the evaluations in the case of 147 research groups in chemistry. The defense at that time was triggered by the introduction of the $h$-index by Hirsch (2005). How did the Leiden "crown" indicators work in comparison to the $h$-index? Unlike the citation indicators, the $h$-index is sensitive to the number of publications for which citation rates are compared. Decomposition of aggregated data allows for distinguishing mechanisms; for example, variance "within groups" versus "between groups."

Since Narin (1976) suggested the use of bibliometrics for evaluative purposes, semi-industrial centers have sprung up either connected to academia (such as in Budapest, Leiden, Leuven, Beijing, Shanghai, etc.) or as independent commercial enterprises (e.g., Science-Metrix in Montreal). Two major companies (Thomson Reuters and Elsevier) are also active in this market. In other words, citation analysis has become an industry. Intellectual property of the data and the results of the analysis has become a major asset in this (quasi-)industry. Although contractors sometimes state that the results are freely available for the users, the licenses of the data (the *Science Citation Index*) often do not permit to publish results freely so that the scientists under study would be able to control these evaluations themselves (cf. Opthof & Leydesdorff, 2010). This practice of secrecy tends to shield the evaluation against the criticism that has been voiced against the use of citation analysis for evaluative purposes (Leydesdorff, 2008; MacRoberts & MacRoberts, 1987, 1996, and 2010).

The invention of the *h*-index as a new statistics in 2005 (Hirsch, 2005), however, challenged one of the leading researchers in the field of evaluative bibliometrics to test whether the new indicator correlated with the "crown" indicators of scientometric evaluation in use by the Leiden Center for Science & Technology Studies (CWTS): CPP/FCSm and CPP/JCSm (Van Raan, 2006). These latter indicators have extensively been used for such purposes as the Leiden Rankings of universities, research evaluation at the institutional level, and science policy advice at national and international (e.g., EU) levels (e.g., Moed, 2005).

The CWTS study (VSNU, 2002) was based on more than 18,000 publications of 147 research groups in chemistry and chemical engineering in the Netherlands for the years 1991-1998. A subset of this data was secondarily analyzed by Van Raan (2006). In addition to the citation indicators, the research groups under study were peer reviewed on their quality on a five-point scale. All fields within chemistry were covered by this set of university groups. The author notes that the various specialties exhibit different citation characteristics and that therefore field-normalization would be essential (cf. Leydesdorff & Opthof, 2010 and 2011). CPP/FCSm normalizes "citations per paper" (CPP) for the mean "field citation score" (FCSm) where a "field" is defined as a set of journals sharing a field-code of the ISI Subject Categories. Analogously CPP/JCSm normalizes for the mean citation scores of individual journals (cf. Waltman *et al.*, 2011b).

Van Raan (2006, at p. 495) provided the following Table 1:[4]

| Research group | P | C | CPP | JCSm | FCSm | CPP/ JCSm | CPP/ FCSm | h-index | Quality |
|---|---|---|---|---|---|---|---|---|---|
| Univ A, 01 | 92 | 554 | 6.02 | 5.76 | 4.33 | 1.05 | 1.39 | 6 | 5 |
| Univ A, 02 | 69 | 536 | 7.77 | 5.12 | 2.98 | 1.52 | 2.61 | 8 | 4 |
| Univ A, 03 | 129 | 3780 | 29.3 | 17.2 | 11.86 | 1.7 | 2.47 | 17 | 5 |
| Univ A, 04 | 80 | 725 | 9.06 | 8.06 | 6.25 | 1.12 | 1.45 | 7 | 4 |
| Univ A, 05 | 188 | 1488 | 7.91 | 8.76 | 5.31 | 0.9 | 1.49 | 11 | 5 |
| Univ A, 06 | 52 | 424 | 8.15 | 6.27 | 3.56 | 1.3 | 2.29 | 9 | 4 |
| Univ A, 07 | 52 | 362 | 6.96 | 4.51 | 5.01 | 1.54 | 1.39 | 8 | 3 |
| Univ A, 08 | 171 | 1646 | 9.63 | 6.45 | 4.36 | 1.49 | 2.21 | 13 | 5 |
| Univ A, 09 | 132 | 2581 | 19.55 | 15.22 | 11.71 | 1.28 | 1.67 | 17 | 4 |
| Univ A, 10 | 119 | 2815 | 23.66 | 22.23 | 14.25 | 1.06 | 1.66 | 17 | 4 |
| Univ A, 11 | 141 | 1630 | 11.56 | 17.83 | 12.3 | 0.65 | 0.94 | 11 | 4 |
| Univ A, 12 | 102 | 1025 | 10.05 | 10.48 | 7.18 | 0.96 | 1.4 | 10 | 5 |

**Table 1:** Example of the results of the bibliometric analysis for the chemistry groups

Table 1 shows the results for 12 research groups in one university who published during this period 1,327 times, obtaining a total of 17,566 citations. The bibliometric indicators, the *h*-index, and the peer ratings are provided. In the latter, "5" indicates "excellent," "4" means "good," and "3" is classified as "satisfactory." Below "3" is not considered "satisfactory," but such a low rating did not occur in this set of data.

---

[4] In footnotes 4 and 5 on p. 464, Van Raan (2006) explains the rationale for using different citation windows for the *h*-index and the CWTS indicators.

|         | CPP/JCSm | CPP/FCSm | h-index | Quality |
|---------|----------|----------|---------|---------|
| *CPP/JCSm* |          | .627*    | .057    | -.230   |
| *CPP/FCSm* | .783**   |          | .352    | .109    |
| *h-index*  | .170     | .219     |         | .169    |
| *Quality*  | -.133    | .156     | .151    |         |

\*\* $p < 0.01$; \* $p < 0.05$

**Table 2:** Pearson correlations (lower triangle) and Spearman rank correlations (upper triangle) among three citation indicators one peer-review based quality indicator

Table 2 shows the Pearson correlations ($r$) in the lower triangle and the Spearman rank correlations ($\rho$) in the upper triangle. As noted (cf. Van Raan, 2006, at p. 499), the *h*-index is also dependent on the number of publications while the CWTS-indicators are not. As could be expected, the two CWTS-indicators are highly correlated between themselves. However, the quality parameter *Q* is uncorrelated with any of these scientometric indicators. Thus, we may conclude that the indicators are *not* validated by this study despite the author's claim to the contrary.

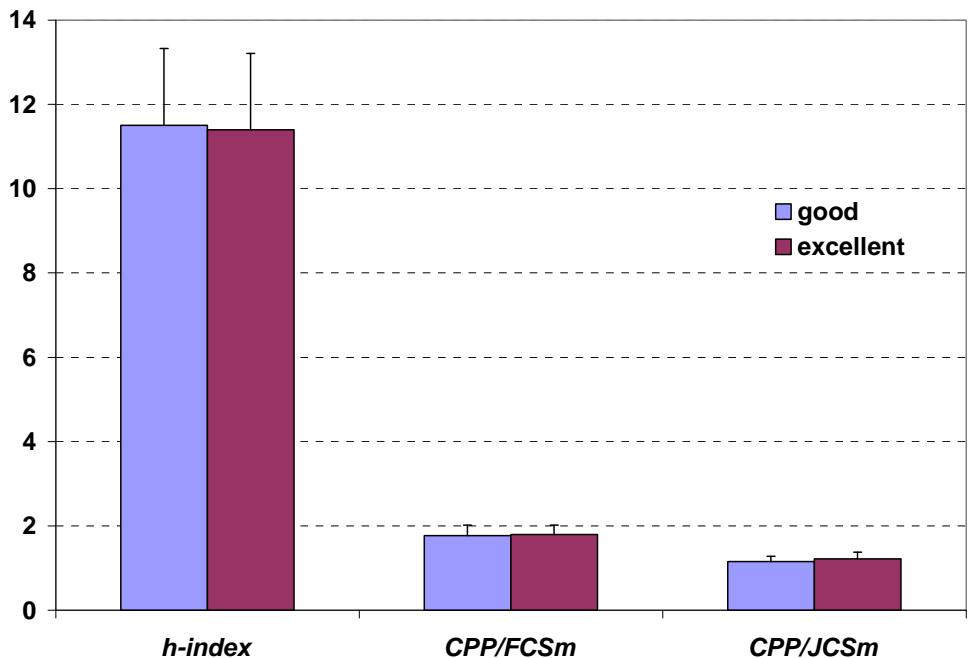

**Figure 1**: Discrimination between "good" and "excellent" research using the *h*-index and the Leiden indicators CPP/JCSm and CPP/FCSm in the case of Table 1.

Figure 1 shows the discriminating power of the *h*-index and the two indicators of CWTS (*CPP/JCSm* and *CPP/FCSm*) using the set provided in Table 1. We added error bars in order to show that the differences are contained within the margins of the standard errors of the measurement. Thus, none of the citation-based indicators is able to discriminate between the categories "good" and "excellent" which were distinguished during the peer review.

In his Table 2, Van Raan (2006, at p. 500) provided also aggregated data for the set of 147 research groups. In this table, the association between *Q* and *h* is significant (using $\chi^2$, and $p < 0.05$), but not the association between *Q* and *CPP/FCSm* when testing $Q = 4$ against $Q = 5$ ($\chi^2 = 4.211$;[5] $df = 2$; $p = 0.112$). Thus, even at this aggregated level ($N = 147$), these results confirm the previous conclusion of Bornmann *et al*. (2010; cf. Van den Besselaar & Leydesdorff, 2009) that the peer review systems and citation analysis are able to distinguish the tails of the distributions (low quality) from the high-end of the set, but perform poorly in distinguishing between excellent and good research to the extent that the relation between qualitative and quantitative impact assessments may be negative (Moed, 2005, at p. 244; Neufeld & Von Iens, in press).

In summary, we argue that the industrial character of citation analysis for evaluative purposes has hidden technical flaws in these measurements because of a lack of openness about the data and therefore critical discussion in academia. Notwithstanding their prevailing use in research evaluation and strategic decision-making, the statistical evidence, for example, supports the claim of the criticizers (e.g., MacRoberts & MacRoberts, 2010) that citation analysis hitherto cannot legitimate the strategic selection of excellence.

---

[5] This value is Yates-corrected because of one value smaller than five. Without this correction: $\chi^2 = 5.559$; $df = 2$; $p = 0.062$.